\documentclass[preprint]{aastex}

\input epsf
\newdimen\fighsize
\def\epsscale#1{\fighsize=#1\hsize}
\def\plotone#1{\centerline{\epsfxsize=\fighsize\hfil\epsfbox{#1}\hfil}}
\def\plottwo#1#2{\centerline{\hbox to \fighsize
                 {\epsfxsize=.5\fighsize\epsfbox{#1}\hfil
                  \epsfxsize=.5\fighsize\epsfbox{#2}}}}

\font\tenib=cmmib10  \skewchar\tenib='177
\font\sevenib=cmmib7 \skewchar\sevenib='177
\font\fiveib=cmmib5  \skewchar\fiveib='177
\def\bmit{\fam15}
\def\vtheta{{\bmit\mathchar"7112}}
\def\vthes{{\vtheta_{\rm s}}}
\def\zl{z_{\rm L}}
\def\Is{I_{\rm s}}
\def\ats{arr\-iv\-al-time surface}

\def\dratio{\Delta t_{2,4}/\Delta t_{1,3}}

\begin{document}

\slugcomment{Version of Apr 19, 01}

\shortauthors{P. Saha \& L.L.R. Williams}

\shorttitle{Einstein Rings}

\title{Beware the Non-uniqueness of Einstein Rings}

\author{Prasenjit Saha}
\affil{Astronomy Unit, School of Mathematical Sciences\\
       Queen Mary and Westfield College\\
       London E1~4NS, UK}
\and
\author{Liliya L.R. Williams}
\affil{Department of Physics and Astronomy\\  
       University of Minnesota\\
       116 Church Street SE\\
       Minneapolis, MN 55455}

\begin{abstract}
We explain how an approximation to the rings formed by the host
galaxies in lensed QSOs can be inferred from the QSO data alone.  A
simple ring image can be made from any lens model by a simple piece of
computer graphics: just plot a contour map of the arrival-time surface
with closely-spaced contours.  We go on to explain that rings should
be (a)~sensitive to time-delay ratios between different pairs of
images, but (b)~very insensitive to $H_0$.  We illustrate this for the
well-known quads 1115+080 and 1608+656.
\end{abstract}

\keywords{gravitational lensing}

\section{Introduction}

It seems likely now that all multiple-image lens systems have Einstein
rings, if imaged deep enough.  Radio rings, formed out of
multiply-imaged radio lobes and jets, have been known for more than a
decade (Hewitt et al.\ 1988, Langston et al.\ 1990, Lehar et al.\ 1993).
But more recently, QSO lenses when
deeply imaged---especially in the infrared---have shown rings formed
out of multiply-imaged parts of the QSO host galaxy: Falco et al.\ (1997)
examined HST images of MG0414+0534 and found an arc connecting three of the
four QSO images; Courbin et al.\ (1997) found an incipient ring in 1115+080; 
Impey et al.\ (1998) uncovered the full ring. 
Blandford, Surpi, \& Kundi\'c (2000, hereafter BSK)
identified ring components in 1608+656. 

Such a rich source of new image data is very exciting, because the
paucity of constraints is a very serious problem in lens modeling,
leading to large uncertainties in current lens reconstructions (see
e.g., Bernstein \& Fischer 1999, Williams \& Saha 2000, hereafter WS).
BSK and Kochanek, Keeton, \& McLeod (2001, hereafter KKM) present new
theoretical ideas for using ring observations to constrain lenses.  In
particular KKM make two very interesting points (see especially their
Figure~1): (i)~in a double (quad), the ring is the image of a two
(four) lobed figure on the source; (ii)~where a ring has brightness
minima, it intersects a critical curve.

This paper presents some further theoretical work, in which we address
the following question: for a lens which {\it already\/} has accurate
image positions and time-delays, how much does a ring {\it further\/}
constrain the lens?  Our approach is very different from previous
ones, and leads to some new insights: the down side is that the same
ring can arise from a large range of lenses and $h$ values, and thus
the prediction by KKM that ``deep, detailed observations of Einstein
rings will be revolutionary for constraining mass models and
determining the Hubble constant from time delay measurements'' is,
unfortunately over-optimistic; the up side is that rings turn out to be
related to time delay measurements in an unexpected but testable way.

\section{Rings and the \ats}\label{sec-two}

While working on WS we noticed that a contour map of the \ats\
in a multiple-image system resembles a ring if the contours
are very close. Figure~14 of WS shows an example.  In that paper we
gave a handwaving argument for why this should happen, but returning
to the question afterwards led to interesting developments, which we
now discuss.

The \ats\ produced by a source at $\vthes$ can be expressed as
\begin{equation}
\tau(\vtheta)=2\,\nabla_\vtheta^{-2}\,\big[1-\kappa(\vtheta)\big]
 - \vtheta\cdot\vthes
\label{eq-tau}
\end{equation}
with $\tau(\vtheta)$ being related to the physical arrival
time\footnote{We will use `\ats' to refer to both $\tau(\vtheta)$ and
$t(\vtheta)$, in places where this causes no ambiguity.}
\begin{equation}
t(\vtheta)=H_0^{-1}(1+\zl){d_{\rm L} d_{\rm S}\over d_{\rm LS}}\,
\tau(\vtheta),
\label{eq-t}
\end{equation}
where the $d$'s are dimensionless angular diameter distances, $\zl$ is
the lens-redshift, and $\kappa(\vtheta)$ the dimensionless mass
profile (see e.g., equations 9 and 10 in Saha 2000).  If the source is
much further than the lens, then
\begin{equation}
t(\vtheta) \simeq h^{-1} \zl(1+\zl) \times 80\,{\rm days\,arcsec^{-2}}\,
\tau(\vtheta).
\end{equation}
If image positions, QSO time delays, redshifts, cosmology, and lens
profile are all known, $H_0$ can be read off from equations
(\ref{eq-tau}) and (\ref{eq-t}).  Unfortunately, this is not the case;
image positions, time delays, and redshifts are accurately measurable,
and cosmology produces a readily-quantifiable (and fairly small)
uncertainty, lens profile is a large uncertainty: a large range of
lens models $\kappa(\vtheta)$ can reproduce QSO image positions and
time delay ratios exactly or nearly exactly, resulting in a range of
plausible $H_0$'s.

An Einstein ring, being an image of an extended source, is not
determined solely by the \ats\ for a point source, but also by the
brightness distribution of the source---see e.g., equation (7) of
KKM. 
KKM and
BSK argue that not all the mass models that reproduce the QSO images
will be able to reproduce the image of the QSO host galaxy.  If so, it
would narrow down the range of acceptable $H_0$'s.

How does an extended source change things?  To study this, let us
consider an extended source with a very simple brightness profile:
peak brightness at $\bar\vthes$ and around it source brightness
falling off in a conical profile
\begin{equation}
\Is(\bar\vthes+\Delta\vthes) = 1 - k\left|\Delta\vthes\right|.
\end{equation}
Here $k$ is a constant, and it is understood that where
$1-k\left|\ldots\right|<0$, the surface brightness is zero.  Say one
of the images of $\bar\vthes$ is at $\bar\vtheta$ on the lens plane,
and that the image-plane region $\bar\vtheta+\Delta\vtheta$ corresponds to
the source-plane region $\bar\vthes+\Delta\vthes$.  Since surface
brightness is preserved, we have for the image brightness
\begin{equation}
I(\bar\vtheta+\Delta\vtheta) = \Is(\bar\vthes+\Delta\vthes)
= 1 - k\left|\Delta\vthes\right|.
\label{eq-I}
\end{equation}
A small displacement in the source plane, $\Delta\vthes$, is related
to the corresponding displacement in the lens plane, $\Delta\vtheta$,
through $\Delta\vthes ={\bf M}^{-1}\cdot\Delta\vtheta$, where ${\bf
M}$ is the inverse magnification. 
With this, the illumination in the lens plane can be rewritten as
\begin{equation}
I(\bar\vtheta+\Delta\vtheta) =
1 - k\left|{\bf M}^{-1}\cdot\Delta\vtheta\right|.
\end{equation}
Now $\tau(\vtheta)$, the arrival time for the point source at
$\vthes$, has zero derivative at $\bar\vtheta$.  In the neighborhood
of $\bar\vthes$, the second derivatives of $\tau$ are simply ${\bf
M^{-1}}$.  Hence in this region, $\nabla\tau$ equals ${\bf
M}^{-1}\cdot\Delta\vtheta$. So the brightness of the ring is
\begin{equation}
I(\vtheta) = 1 - k\left|\nabla\tau\right|.
\label{eq-ring}
\end{equation}
Now picture a sheet of paper with contours of arrival time drawn on
it, viewed so that nearby contour lines blur into a
grayscale. Equation~\ref{eq-ring} then also happens to represent the
grayscale pattern on this sheet of paper, where
\begin{equation}
k \propto {\langle\hbox{thickness of contour lines}\rangle \over
           \langle\hbox{$\tau$-spacing between contour lines}\rangle},
\label{eq-k}
\end{equation}
with `1' representing blank paper, and $k\left|\nabla\tau\right|$ the
darkening due to the contour lines.  We thus have a quantitative
explanation for why Figure~14 of WS looks like a ring.

More importantly, we see that given a lens model for QSO images,
generating a ring image for a conical source is very simple: one has
only to plot a closely spaced contour map of the \ats; the only
adjustable parameter is the contour interval, which behaves as a
surrogate for the source size $k^{-1}$.

The relation (\ref{eq-ring}) between the \ats\ (of the QSO or other
compact source) and the ring (formed from the host galaxy) provides
two insights, which are quite generic and not limited to conical
sources.
\begin{enumerate}
\item Consider time delays versus image separations between different
pairs of images in a quad.  Where the ratio of time delay to image
separation is small, the density of contour lines will be small,
therefore the corresponding segment of the ring will be brighter.
Thus, if time delay ratios in a quad have been measured then the
relative brightness of different ring segments can be predicted to
some extent.  Conversely, if QSO images and the image of the ring are
observed then time delay ratios can be estimated.  
\item Altering the lens mass model, and hence $\tau(\vtheta)$ in such
a way that there is little or no change in the density of contour
lines in the region of the ring will produce little or no change in
the ring (other than its overall brightness), but may cause a large
change in $H_0$.  One example is evident from equations (\ref{eq-tau})
and (\ref{eq-t}): if $[1-\kappa(\vtheta)]$, $\vthes$, and $H_0$ are
all multiplied by the same constant factor, $t(\vtheta)$ is
unaffected.  This is the mass-disk degeneracy, first noted by Falco et
al.\ (1985) and discussed in several recent papers, e.g., Saha (2000);
basically it means that steeper lens profile give higher $H_0$ from
the same data---see the ``radial-index versus $h$'' plots in WS for
examples. This particular degeneracy is exact, but numerically one can
cook up any number of near-degeneracies that scale $H_0$ significantly
while producing only slight, observationally negligible, changes in
rings.
\end{enumerate}

In the following section we follow up the implication of these two
points in detail for two quads.

\section{Case studies: 1115+080 and 1608+656}

These are among the most-studied lenses: 1115+080 was discovered by
Weymann et al.\ (1980) and its time delays were determined by
Schechter et al.\ (1997); 1608+656 was discovered by Myers et al.\
(1995) and Fassnacht et al.\ (1999) measured its time delays.

\begin{figure}
\epsscale{.55}
\plottwo{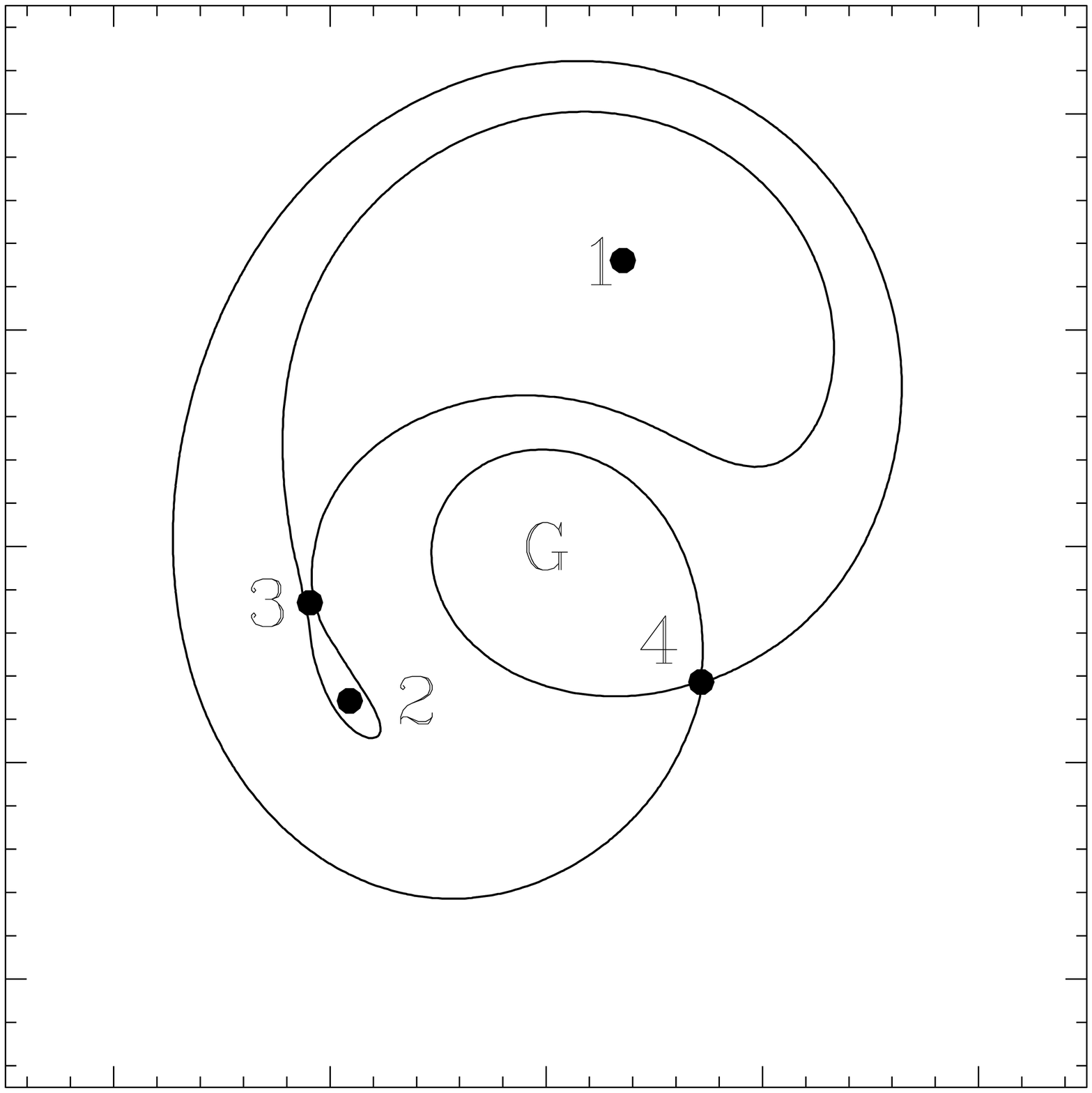}{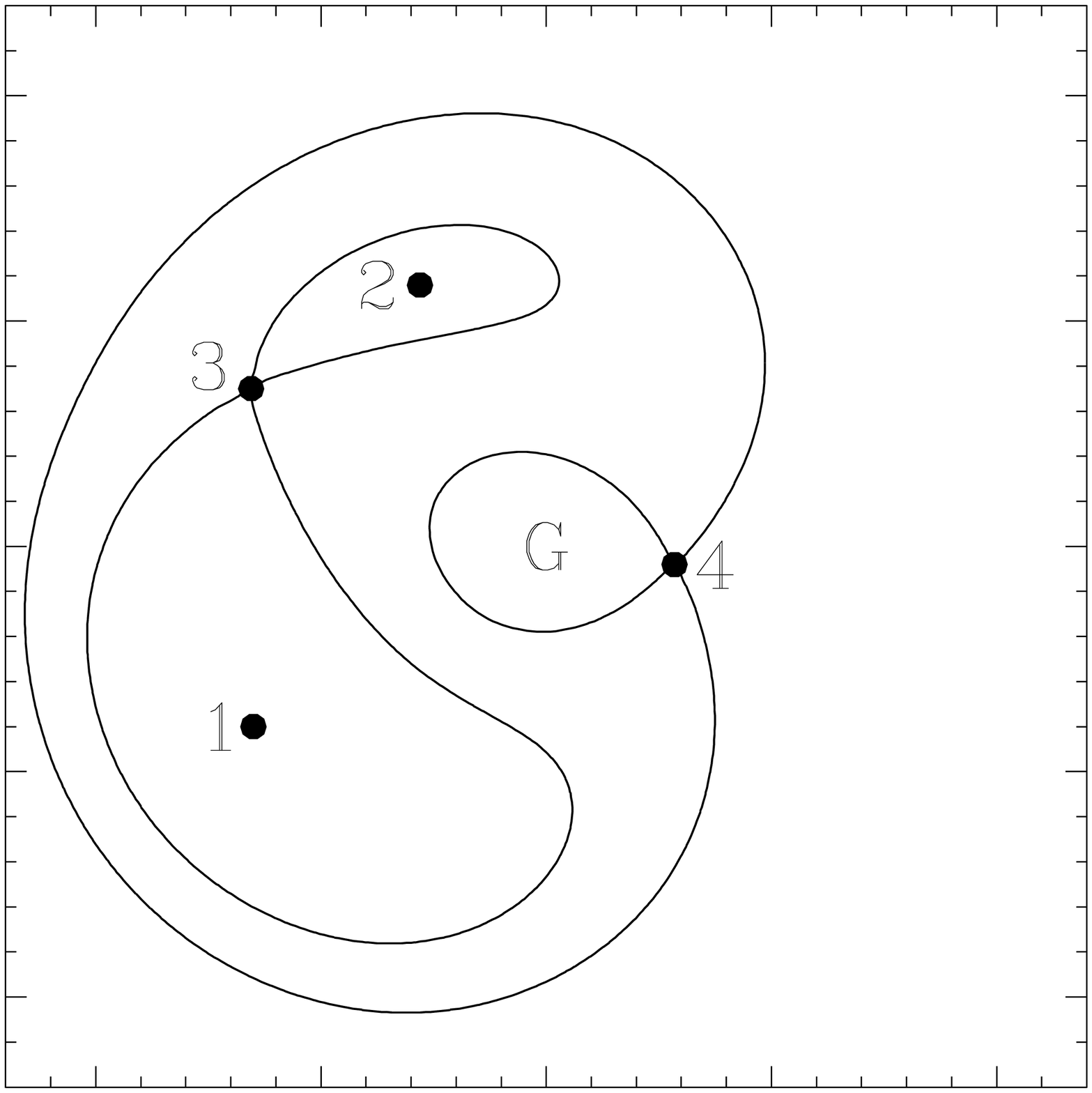}
\caption{Image configuration and possible saddle-point contours on the
\ats\ for 1115+080 and 1608+656.  Images are labelled by increasing
arrival time, and G labels the center of the main lensing galaxy.}
\label{fig-sad}
\end{figure}

Figure \ref{fig-sad} shows the image configurations for both lenses.
We have plotted the saddle-point contours, which make up a sort of
skeleton for the \ats, and renamed the images by arrival time.  Upon
doing this, it becomes evident how similar these two lenses are.

We can now use the ideas of the previous section to predict the
morphology of the rings.  The contour-density argument show that
\begin{equation}
    {{(\Delta\tau_{1-3}/\Delta\theta_{1-3})}
\over{(\Delta\tau_{2-4}/\Delta\theta_{2-4})}}
\approx
    {\hbox{brightness of 2--4 ring segment}
\over\hbox{brightness of 1--3 ring segment}},
\label{eq-segment}
\end{equation}
and similarly for any pair of ring segments.  Now, arrival-time
contours will be furthest apart in the region 2--3, closer together in
1--3 and 2--4, and closest in 1--4. So the brightest part of the ring
will be 2--3, while 1--4 will be a faint arc or a gap in the ring.  The
same conclusions follow from a more complicated argument based on
caustics in Blandford \& Kovner (1988).  And they may be verified by
inspecting Figure~1 of Impey et al.\ (1998) and Figure~1 of BSK.

We now show the predicted rings from models that precisely fit the QSO
image and time-delay data.  We generate ensembles of such models using
the method developed in WS, with one change: instead of letting $h$
vary in the ensemble and then examining its range, we fix it at some
trial value; in this way we can examine how the predicted ring varies
with the trial $h$.  All models we show in this paper are averages of
ensembles of 100.  We have not attempted to fit for the contour step,
we have just chosen a suitable value empirically.

\begin{figure}
\epsscale{0.27}
\plotone{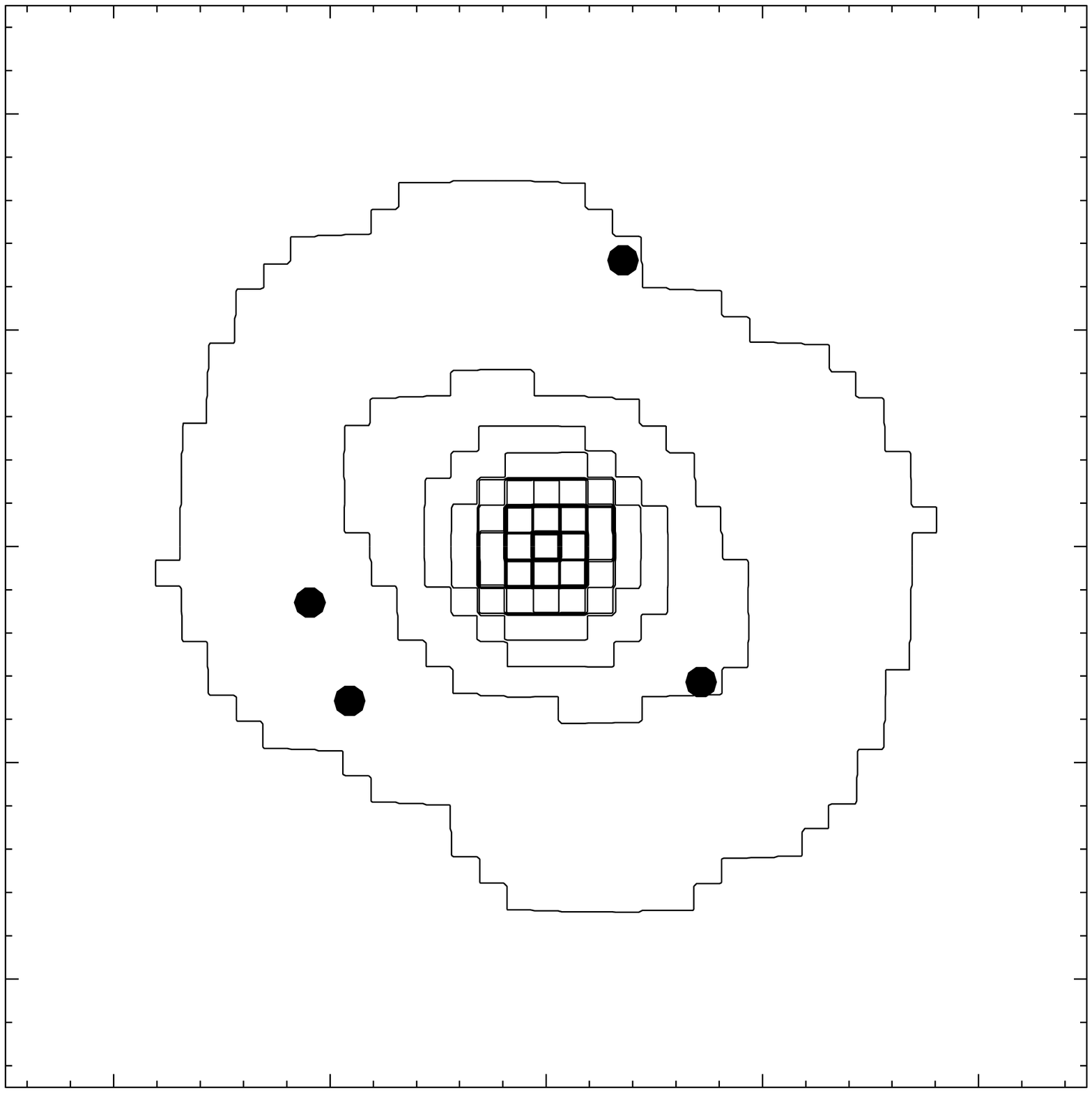}
\epsscale{0.55}
\plotone{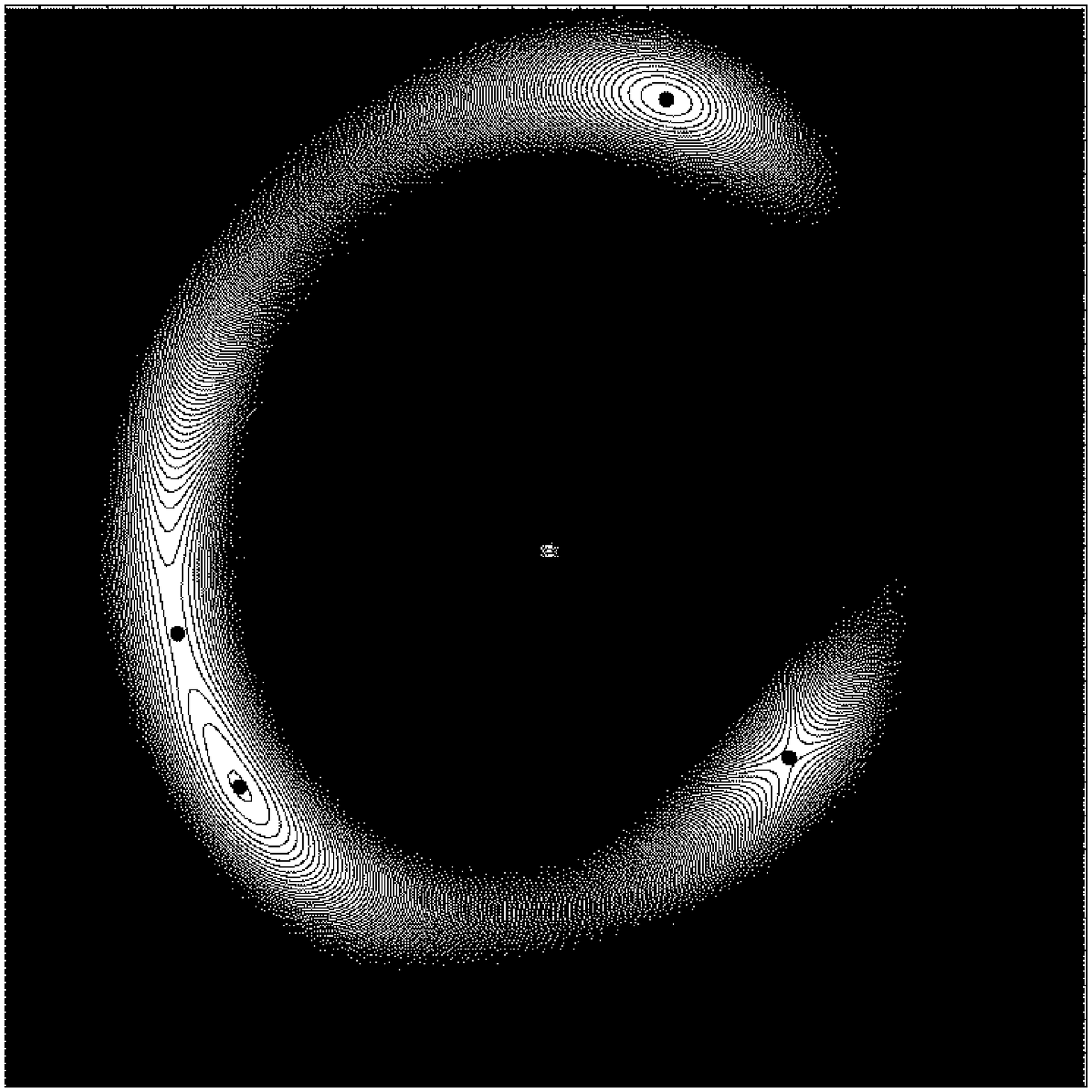}
\caption{An ensemble-average model of 1115+080, constrained to give
$h=0.5$ for the Barkana (1997) estimates for the time delays. In
particular the time delay ratio is $\dratio=0.88$. The upper panel is
the mass map, with $\kappa$ in steps of $1\over3$.  The lower panel
shows the \ats\ with contours in spaced by 2\thinspace
hr; this may be compared with the lower-left panel in Figure~1 of Impey
et al.\ (1998).}
\label{fig-bk50}
\end{figure}

\begin{figure}
\epsscale{0.27}
\plotone{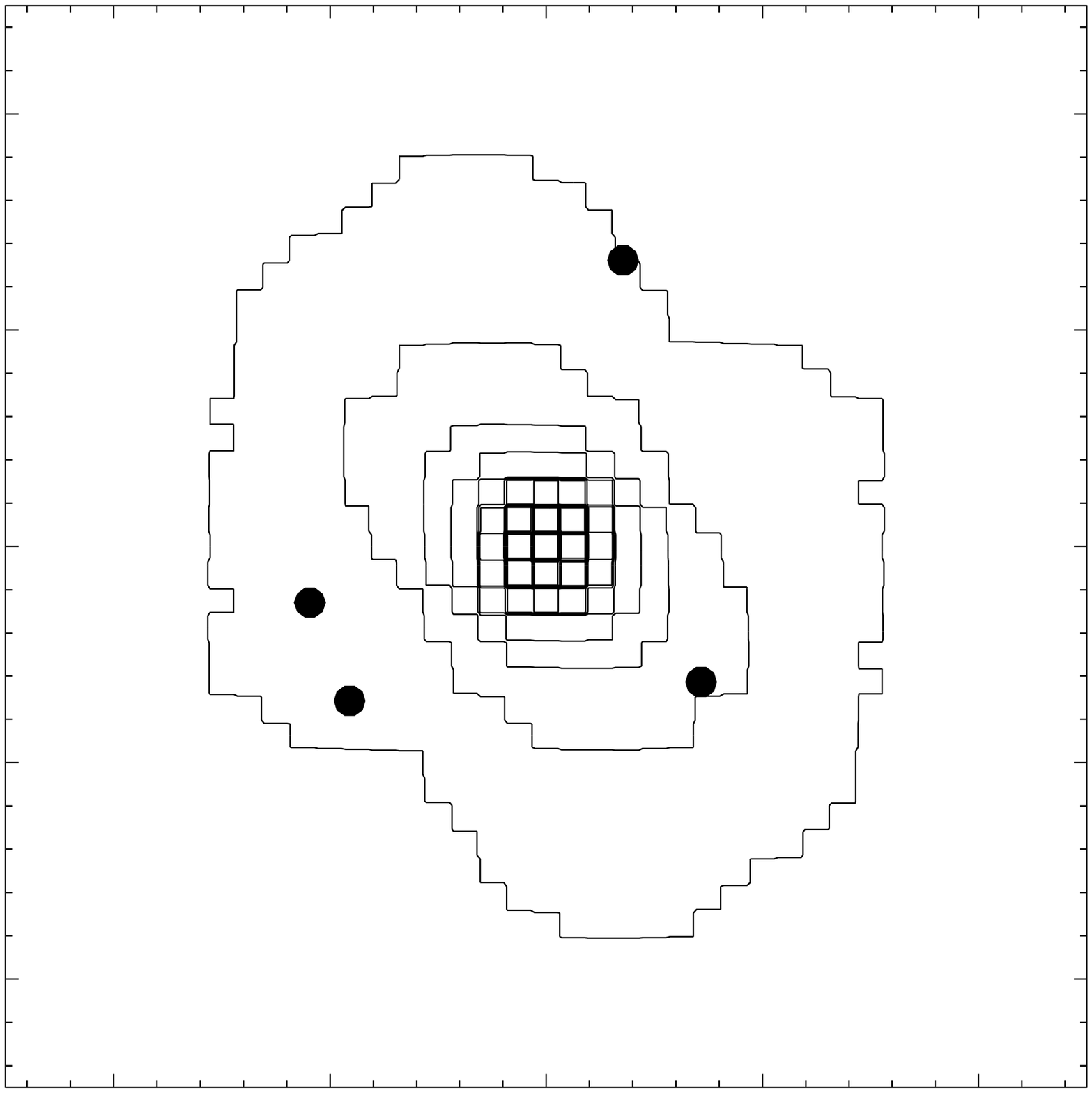}
\epsscale{0.55}
\plotone{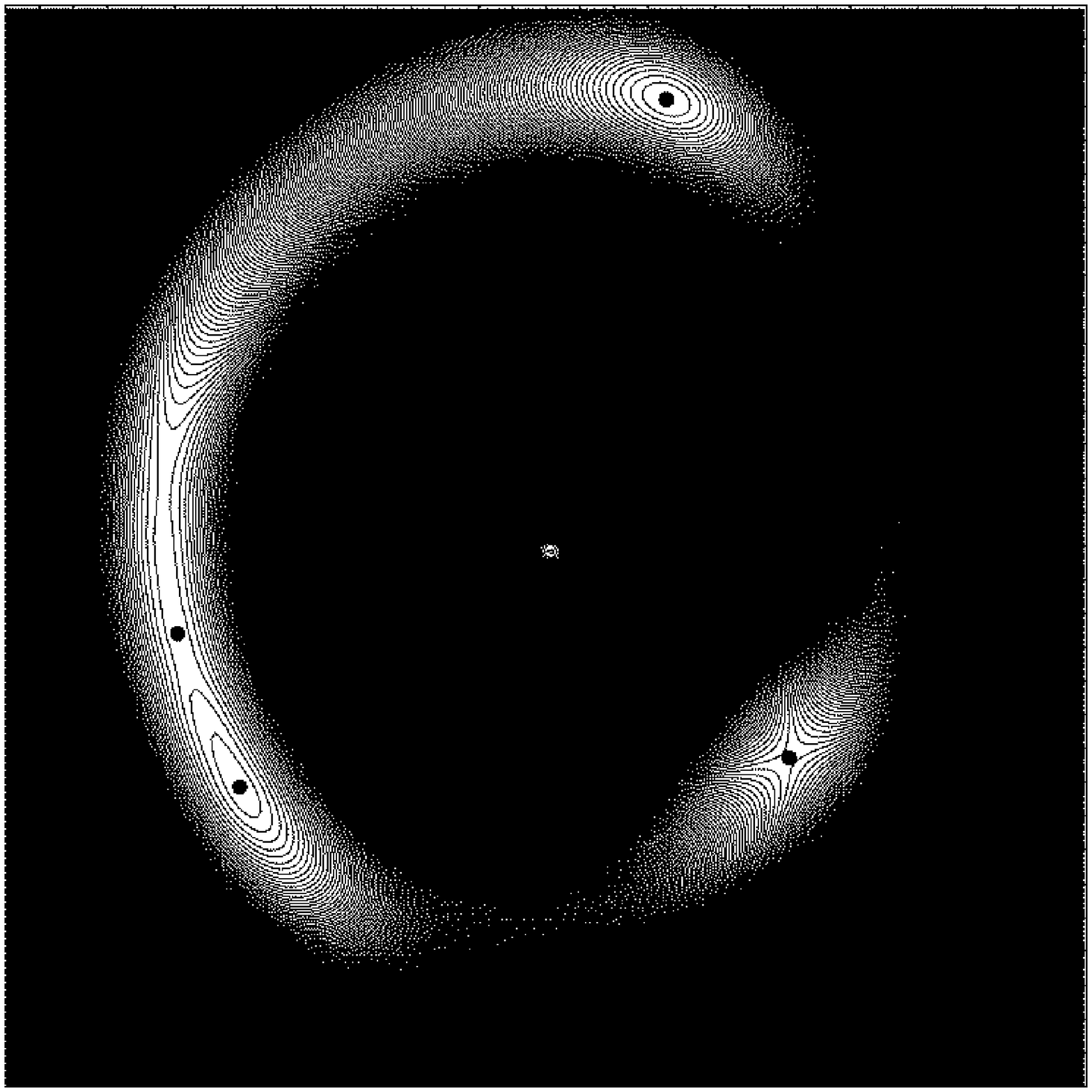}
\caption{An ensemble-average model of 1115+080 constrained to give
$h=0.5$ from the Schechter et al.\ (1997) estimates for the time
delays, in particular $\dratio=1.67$. Note how increasing $\dratio$
has made the 1--3 section of the ring brighter and and the 2--4
section fainter.}
\label{fig-s50}
\end{figure}

Figure \ref{fig-bk50} shows a model of PG1115, for $h$ fixed at 0.5.
The model is constrained by image positions and Barkana's (1997)
estimate of the time delay from the Schechter et al.\ (1997) data.  It
predicts the gross features of the observed ring.  Figure
\ref{fig-s50} is similar ($h$ is also set to 0.5), but uses the
earlier preliminary time delay estimate by Schechter et al.\ (1997);
the only significant difference is that $\dratio$ is larger, and in
response the ring segment 2--4 has become fainter than 1--3.  In the
observed ring the 2--4 segment is actually slightly brighter than
1--3.  (See Figure~1 of Impey et al.\ 1998, and also Figure 3 of KKM).
Thus, the observed ring favors the Barkana estimate of $\dratio$.
Moreover, since the model is still slightly under-predicting the
brightness of 2--4 versus 1--3, we further predict that as the
time-delay measurements improve, $\dratio$ will get somewhat smaller.
This illustrates point 1 from the end of Section~\ref{sec-two}, that
QSO image positions, time delay ratios, and the ring are not all
independent.

\begin{figure}
\epsscale{0.54}
\plottwo{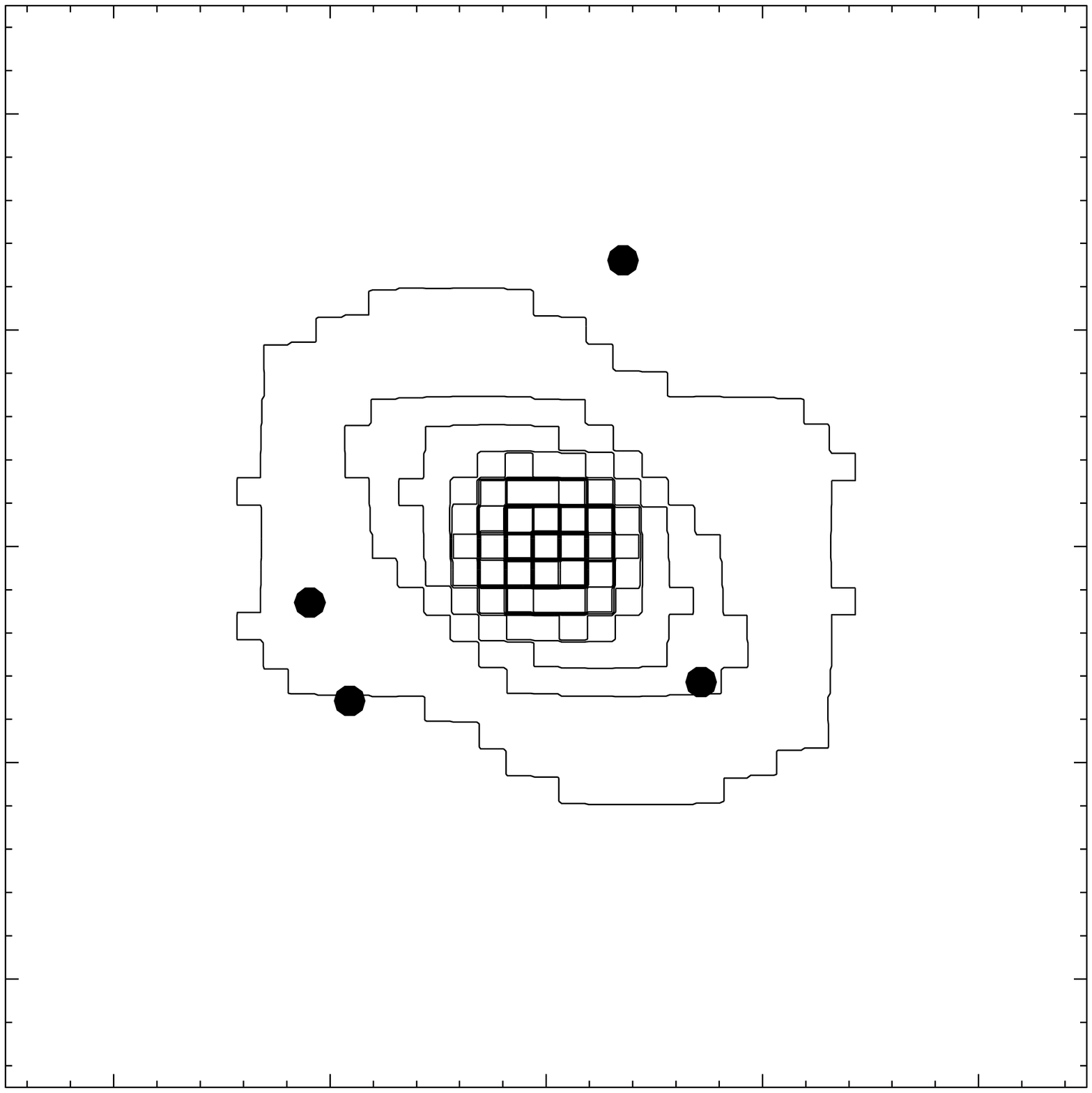}{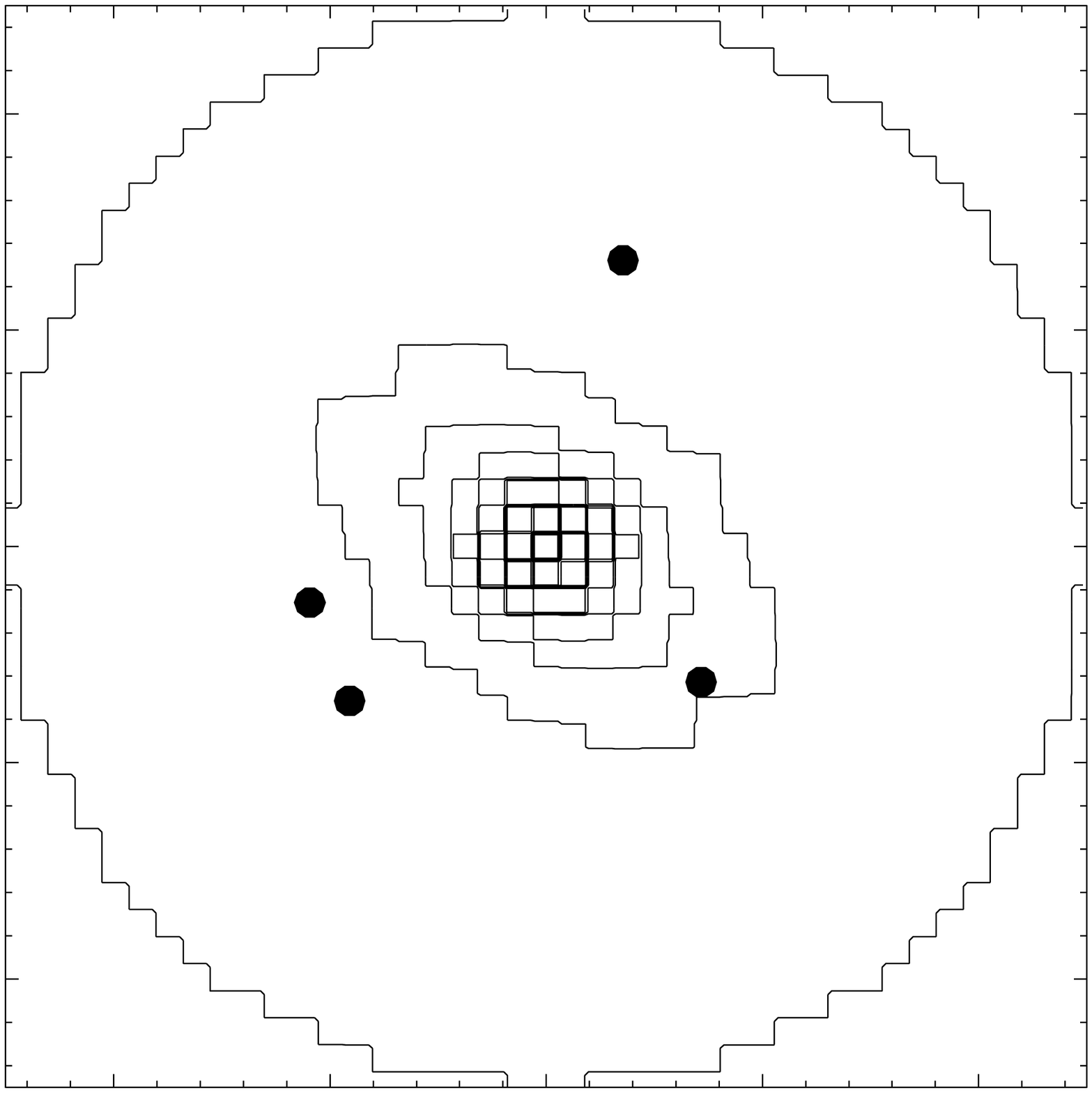}
\epsscale{0.55}
\plotone{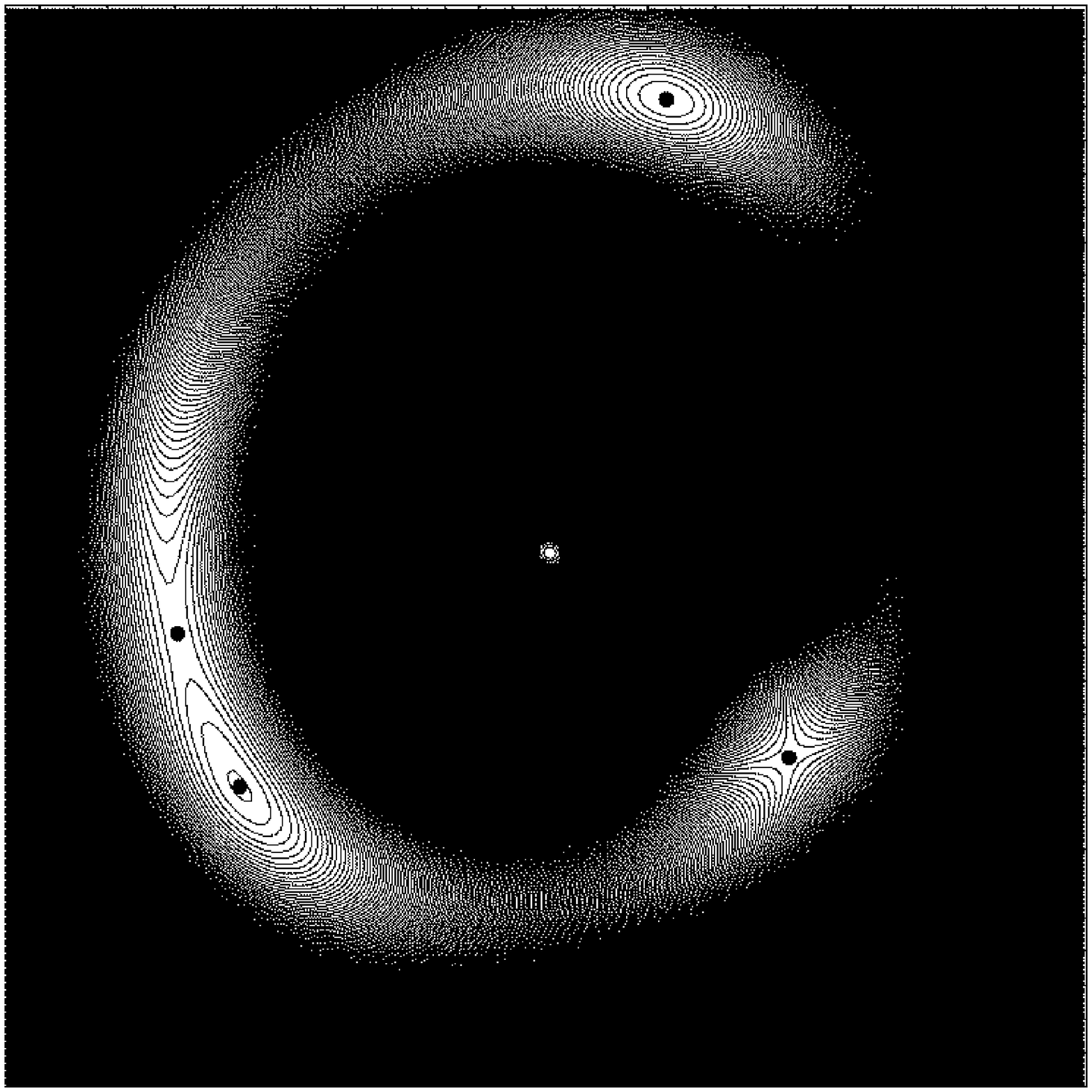}
\caption{Another ensemble-average models of 1115+080, this time
constrained to give $h=0.75$ from the Barkana (1997) estimates for the
time delays.  The mass map is on the upper left; the upper right panel
shows the mass map transformed using the mass disk degeneracy to give
the same time delays for $h=0.5$---note that the critical-density
contour (third from the outside) does not move.  The $t(\vtheta)$
shown is identical for both mass maps.}
\label{fig-bk75}
\end{figure}

The above conclusions do not depend on the trial value chosen for $h$.
Figure~\ref{fig-bk75} shows a model using the Barkana values and
$h=0.75$.  The ring is similar to that in Figure~\ref{fig-bk50}
(Barkana values and $h=0.5$), and demonstrates that QSO image
positions and time delays determine the ring, while the mass
distribution can vary substantially: compare the top right mass map in
Figure~\ref{fig-bk75} to that in Figure~\ref{fig-bk50}.  This
illustrates point 2 from the end of Section~\ref{sec-two}.  It is also
contrary to the assertion by KKM that ``The Einstein ring in
PG1115+080 rules out the centrally concentrated mass distributions
leading to a high Hubble constant ($H_0>60\rm\,km\,s^{-1}\,Mpc$) given
the measured time delays''.\footnote{KKM's limit actually uses the
preliminary Schechter et al.\ value for $\dratio$, rather than the later
Barkana value.  Our argument would not change if we used the other
value.}  Incidentally, note that the $h=0.75$ model in
Figure~\ref{fig-bk75} is much more elliptical than the $h=0.5$ model
in Figure~\ref{fig-bk50}, but it is not much more centrally
concentrated.  It is easier to produce $h>0.6$ models for this system
if they are more centrally concentrated, but it is not necessary (see
Figure~13 of WS).

\begin{figure}
\epsscale{0.27}
\plotone{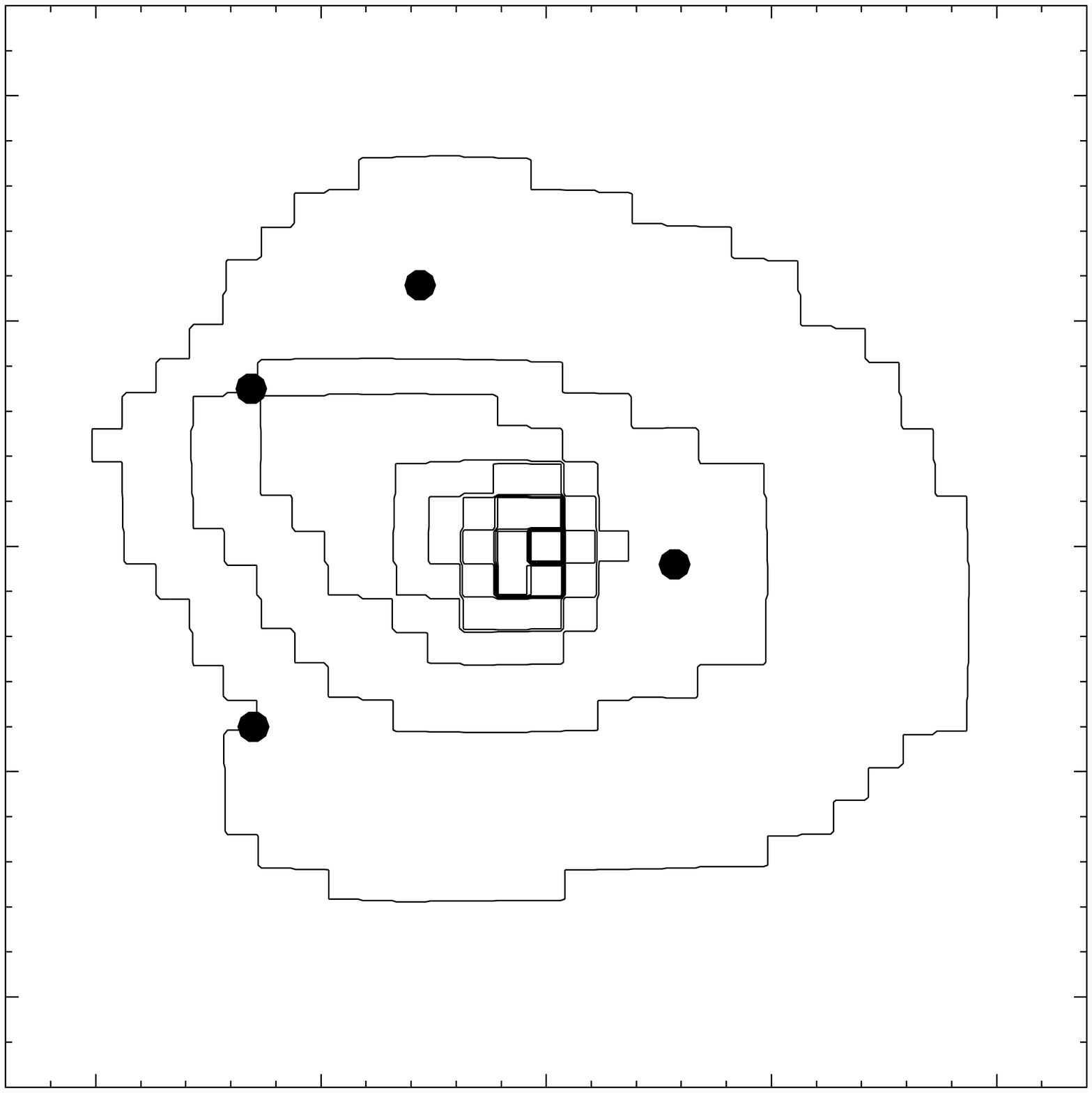}
\epsscale{0.55}
\plotone{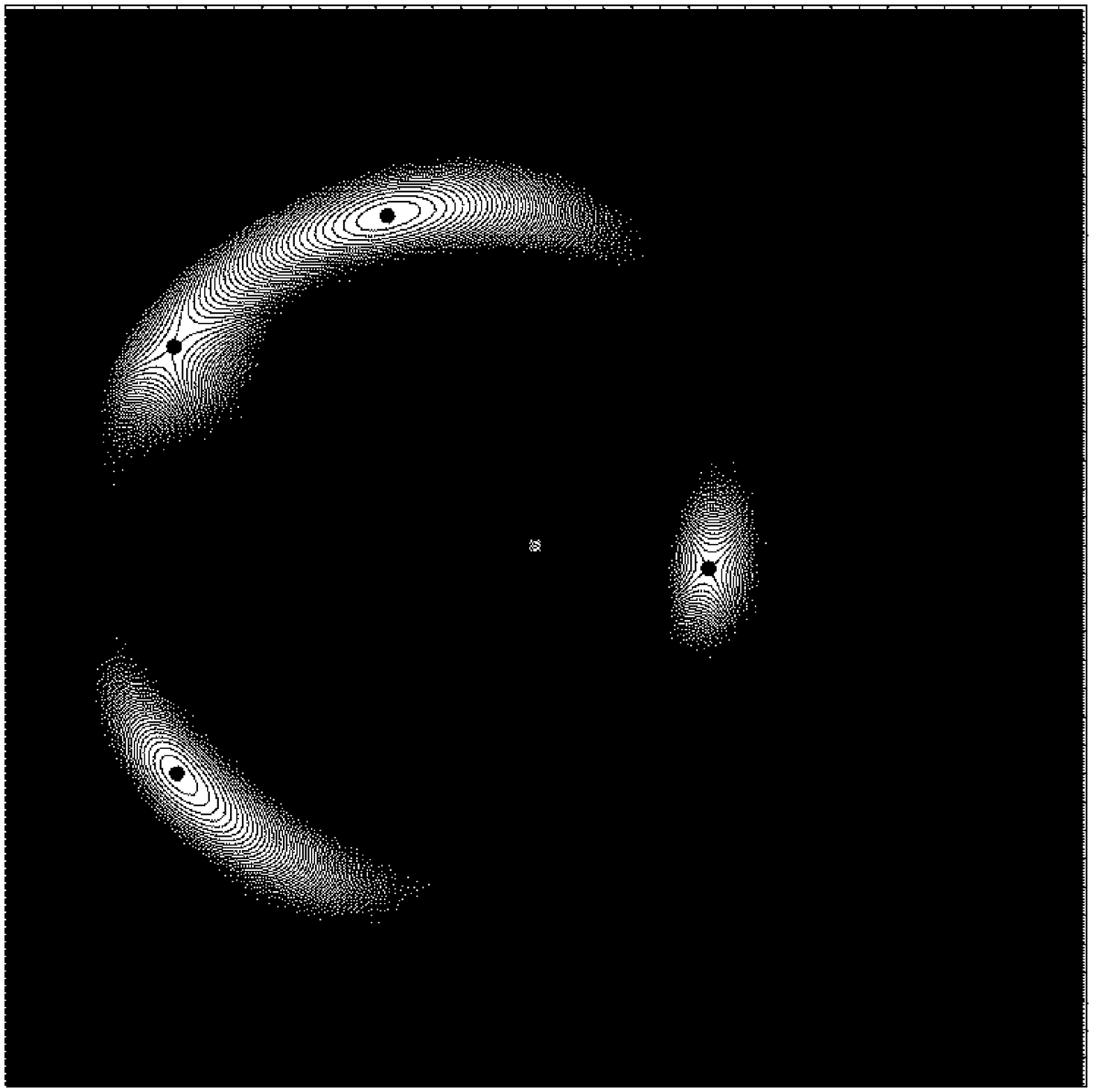}
\caption{An ensemble-average model of 1608+686, constrained to give
$h=0.5$ for the Fassnacht et al.\ (1999) time delays.  The
arrival-time contours are spaced by 6\thinspace hr. The lower panel
here, after rotating clockwise by $90^\circ$, may be compared with the
lower-left panel in Figure~1 of BSK}
\label{fig-f50}
\end{figure}

\begin{figure}
\epsscale{0.54}
\plottwo{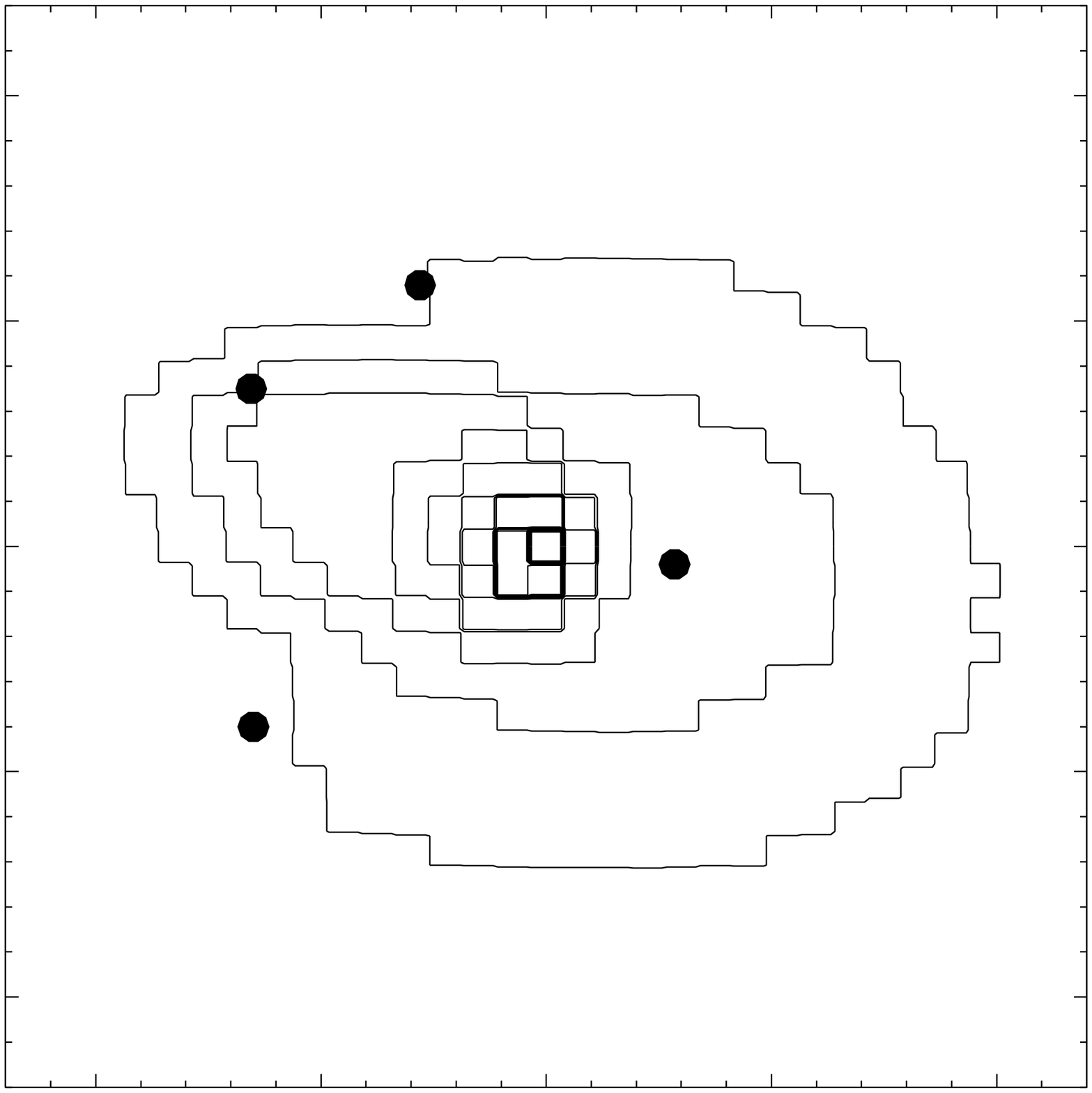}{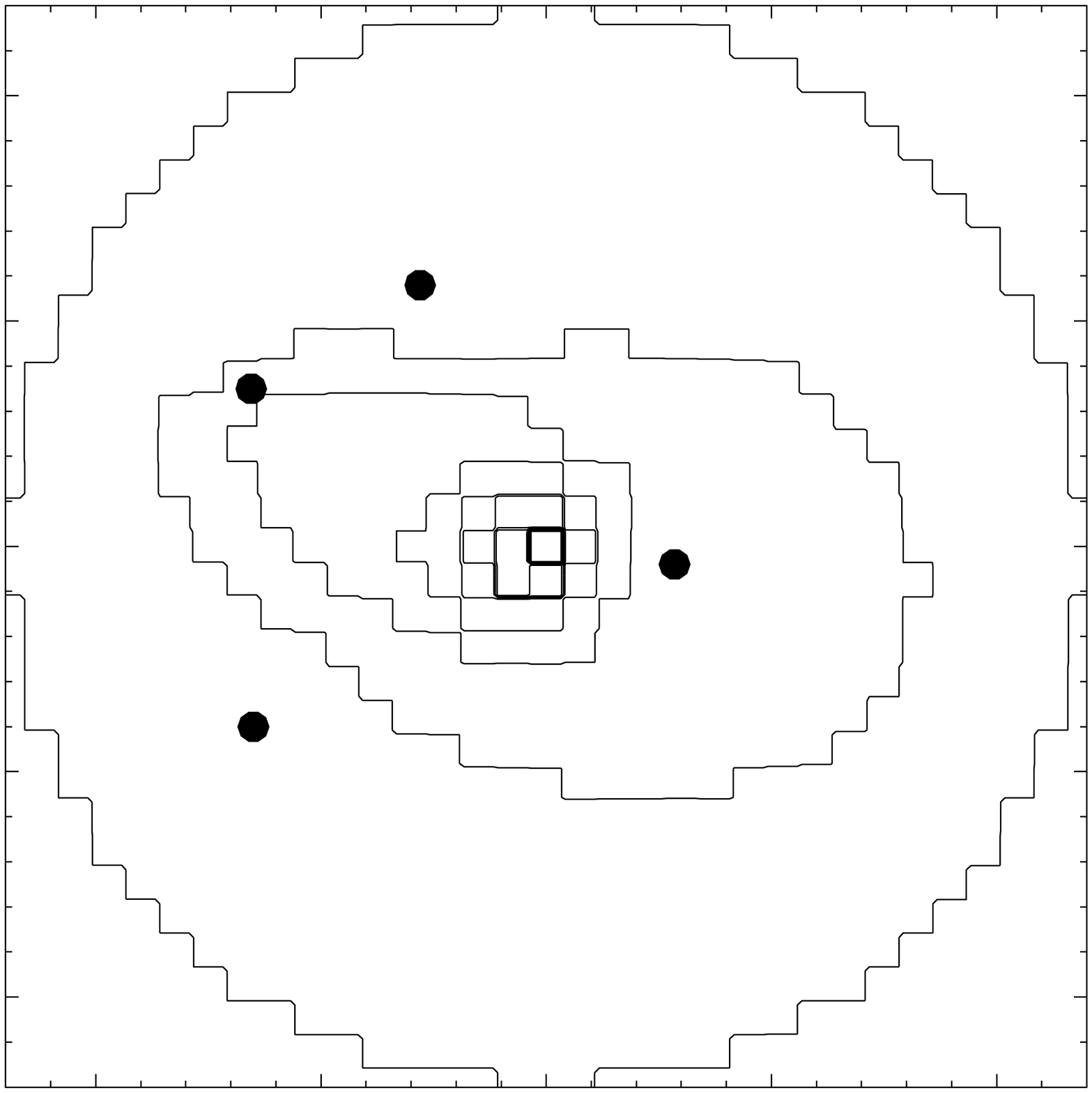}
\epsscale{0.55}
\plotone{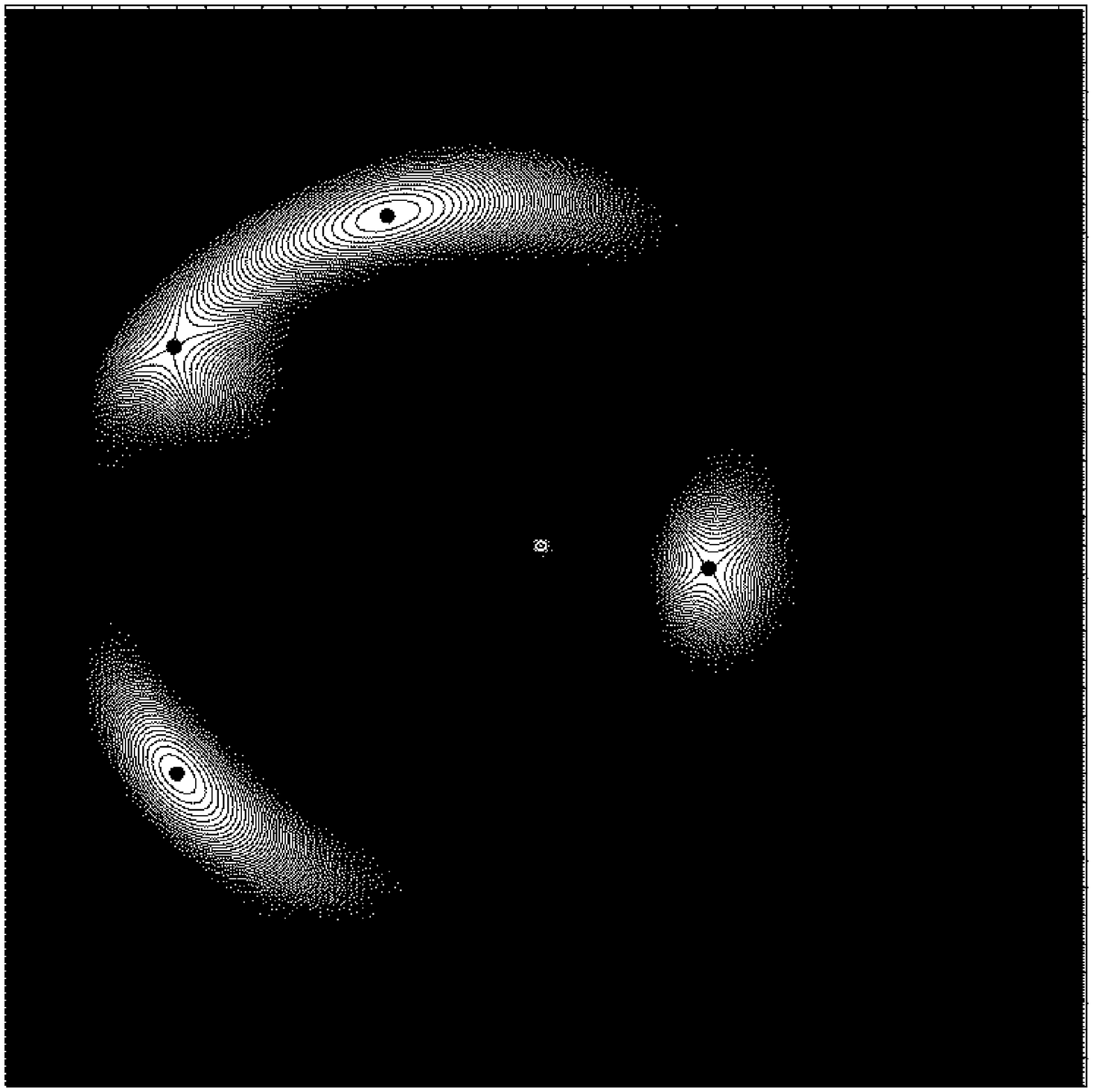}
\caption{An ensemble-average model of 1608+686, constrained to give
$h=0.75$ for the Fassnacht et al.\ (1999) time delays.  The mass map
is on the upper left; the upper right panel shows the mass map
transformed using the mass disk degeneracy to give the same time
delays for $h=0.5$.  The ring applies to either mass map.}
\label{fig-f75}
\end{figure}

Figures \ref{fig-f50} and \ref{fig-f75} are for 1608+656, and are
analogous to Figures~\ref{fig-bk50} and \ref{fig-bk75}.  The data are
from Fassnacht et al.\ (1999) and the predicted rings may be compared
with Figure~1 of BSK.  Again, we find that changing $h$ makes little
difference to the ring.

Why do we, though we have no disagreement with the basic theoretical
ideas in KKM, come to the opposite conclusion regarding constraints on
$h$?  The reason lies in the modeling strategy.  KKM assume a fixed
functional form for the lens described by a few parameters; these
models have little freedom to vary radial profile, and do not allow
ellipticity to vary or to twist with radius.  Pixelated models, such
as the ones in this paper, do not have such restrictions and thus are
able to expose the degeneracies left by rings.  Of course it is not
necessary to use pixelated models for this purpose, parametric models
with some additional parameters would also be effective.  What these
additional parameters need to be is not obvious, but from examining
the pixelated mass maps we guess that letting the index of the radial
profile vary and letting the ellipticity vary with radius are probably
the most important ingredients.

\section{Conclusions}

We show that the \ats\ for a multiply-imaged QSO (or other compact
source), when plotted as a contour map with densely packed contours,
is approximately the lensed image of an extended source with a conical
surface-brightness profile.  An observed ring will depend on the
source's detailed brightness profile, so our result does not furnish a
method for making detailed models of rings.  However, our result does
imply that the gross features of a ring are determined by the
information encoded in the QSO point images alone, which leads to two
useful findings, as follows.
\begin{enumerate}
\item It becomes possible to translate statements about time delays
into statements about the ring (eq.~\ref{eq-segment}).  
In particular, the general morphology
of rings in quads becomes very easy to predict, from just the
configuration of the QSO images. For the case of 1115+080, the ring
can even be used to predict the sign of the current errors in the
ratio of two time delays.
\item It becomes evident that the same ring can arise from a large
range of mass distributions, and hence from a large range of $h$.  The
main (but not the only) contributor to this non-uniqueness is the mass
disk degeneracy, which makes the lens mass profile steeper while
making the Universe smaller but keeps image and time-delay data the
same.  We illustrate non-uniqueness in models of 1115+080 and
1608+656.
\end{enumerate}

\acknowledgments We thank J.C.B. Papaloizou for elucidating the
connection between contour maps and ring images.

\end{document}